\documentclass{vgtc}                          % final (conference style)
\ifpdf%                                % if we use pdflatex
  \pdfoutput=1\relax                   % create PDFs from pdfLaTeX
  \usepackage{graphicx}                % allow us to embed graphics files
  \DeclareGraphicsExtensions{.pdf,.png,.jpg,.jpeg,.JPG,.PNG} % for pdflatex we expect .pdf, .png, or .jpg files
\else%                                 % else we use pure latex
  \usepackage{graphicx}                % allow us to embed graphics files
  \DeclareGraphicsExtensions{.eps}     % for pure latex we expect eps files
\fi%

\graphicspath{{figures/}{pictures/}{images/}{./}} % where to search for the images

\usepackage{microtype}                 % use micro-typography (slightly more compact, better to read)
\PassOptionsToPackage{warn}{textcomp}  % to address font issues with \textrightarrow
\usepackage{textcomp}                  % use better special symbols
\usepackage{mathptmx}                  % use matching math font
\usepackage{times}                     % we use Times as the main font
\usepackage{cite}                      % needed to automatically sort the references
\usepackage{tabu}                      % only used for the table example
\usepackage{booktabs}                  % only used for the table example
\usepackage{amsmath}

\onlineid{1002}

\vgtccategory{Research}

\vgtcinsertpkg

\newcommand{\etal}{\emph{et al.}}

\title{Inverse Augmented Reality: A Virtual Agent's Perspective}

\author{Zhenliang Zhang\\
    \scriptsize Beijing Institute of Technology
\and Dongdong Weng\thanks{e-mail: crgj@bit.edu.cn}\\ 
\parbox{1.4in}{\scriptsize \centering Beijing Institute of Technology \\ AICFVE of Beijing Film Academy}
\and Haiyan Jiang\\ 
    \scriptsize Beijing Institute of Technology
\and Yue Liu\\  
    \parbox{1.4in}{\scriptsize \centering Beijing Institute of Technology \\ AICFVE of Beijing Film Academy}
\and Yongtian Wang\\  
    \parbox{1.4in}{\scriptsize \centering Beijing Institute of Technology \\ AICFVE of Beijing Film Academy}
     }

\author{
Zhenliang Zhang$^{1}$, Dongdong Weng$^{1,2,}$\thanks{crgj@bit.edu.cn},
 Haiyan Jiang$^{1}$, Yue Liu$^{1,2}$, Yongtian Wang$^{1,2}$ \\
    \scriptsize 
    \leftline{
    \lbrack 1\rbrack\ Beijing Engineering Research Center of Mixed Reality and Advanced Display, Beijing Institute of Technology, Beijing, China}\\ 
    \scriptsize 
    \leftline{
    \lbrack 2\rbrack\ AICFVE of Beijing Film Academy, Beijing, China}
    }

\abstract{We propose a framework called inverse augmented reality (IAR) which describes the scenario that a virtual agent living in the virtual world can observe both virtual objects and real objects. This is different from the traditional augmented reality. The traditional virtual reality, mixed reality and augmented reality are all generated for humans, i.e., they are human-centered frameworks. On the contrary, the proposed inverse augmented reality is a virtual agent-centered framework, which represents and analyzes the reality from a virtual agent's perspective. In this paper, we elaborate the framework of inverse augmented reality to argue the equivalence of the virtual world and the physical world regarding the whole physical structure.  
} % end of abstract

\CCScatlist{
  \CCScatTwelve{Human-centered computing}{Human computer interaction}{Interaction paradigms}{Mixed / augmented reality};
  \CCScatTwelve{Computing methodologies}{Computer graphics}{Graphics systems and interfaces}{Mixed / augmented reality}
}

\begin{document}

%% The ``\maketitle'' command must be the first command after the
%% ``\begin{document}'' command. It prepares and prints the title block.

%% the only exception to this rule is the \firstsection command
\firstsection{Introduction}

\maketitle
The basic framework for augmented reality (AR), mixed reality (MR) and virtual reality (VR) was proposed by Milgram \etal \cite{milgram1994taxonomy}. These paradigms are designed for the human-centered world. As the artificial intelligence develops rapidly, a virtual agent will finally possess an independent mind similar to that of humans. Based on Minsky's analysis of the human's mind \cite{minsky1988society}, a virtual agent could develop its own independent mind and live successfully in the virtual world as humans can do in the real world. For this reason, a virtual agent can have an equal status with real humans. The well-known augmented reality can transfer from a human-centered framework to a virtual agent-centered framework. When the virtual agent is the center of the system, it can observe both virtual objects in the virtual world and real objects in the real world. This is called inverse augmented reality (IAR), because it uses an exactly opposite observing direction compared to the traditional augmented reality.

The idea of IAR is originally inspired by the concept of the parallel world in the discipline of physics \cite{berezhiani2004mirror}. Based on the consideration of physics, IAR requires that the virtual world exists with similar structures and interaction roles to that of the physical world. These similar structures and interaction roles have been applied to virtual reality in order to define inverse virtual reality (IVR) \cite{zhang2018Inverse}. In this paper, we would talk about inverse augmented reality using the similar methodology.

The study about IAR is significant for two following reasons. First, it figures out the relationship between the virtual world and the physical world under the background of IAR, promoting the development of the scientific architecture of virtual agent-centered inverse augmented reality. Second, it lays the foundation of inverse augmented reality applications which do not treat the human as the system center, increasing the diversity of augmented reality systems. For these reasons, the proposed IAR is expected to make a breakthrough in both theory and practice.

\begin{figure}[t]
 \centering % avoid the use of \begin{center}...\end{center} and use \centering instead (more compact)
 \includegraphics[width=\columnwidth]{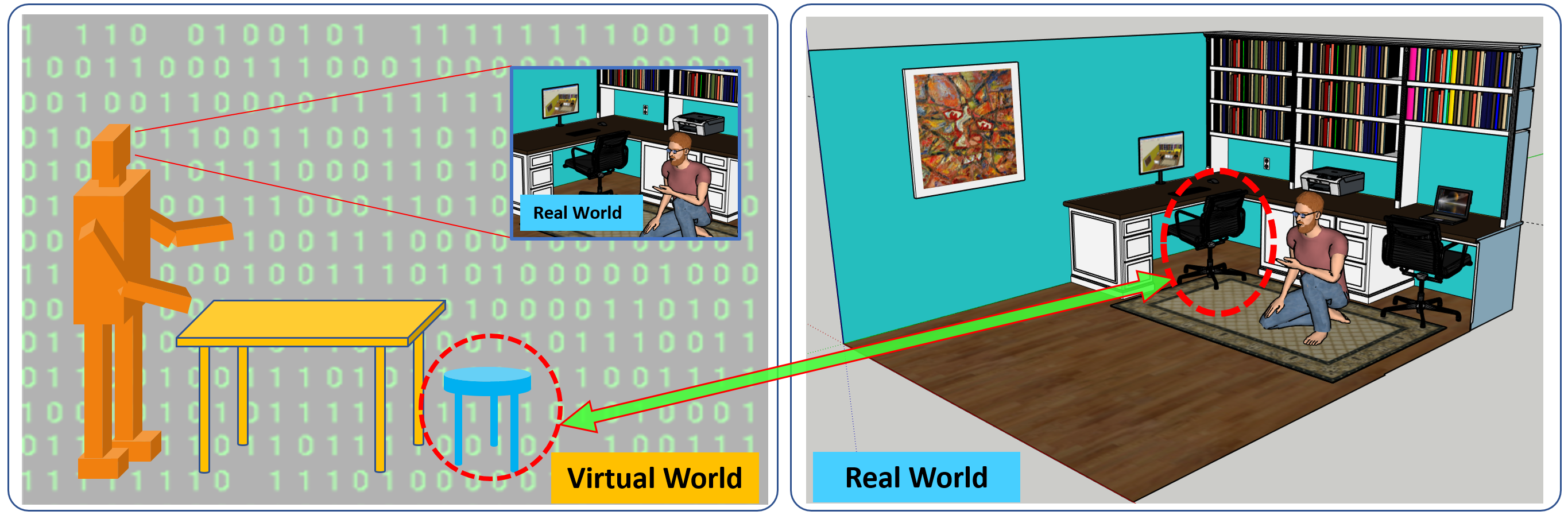}
 \caption{A typical scene of inverse augmented reality. In the left side, the virtual agent is represented as an orange avatar. A real chair is registered into the virtual world, so that a virtual one corresponds to a real one. Meanwhile, the virtual yellow table in the virtual world can exist independently with no relationship with the desks in the real world. The real world can be observed by the virtual agent, but only the registered real objects are available data which can augment the virtual world.  }
 \label{fig:demo}
\end{figure}

This paper proposes the concept of IAR, and concretely shows the relationship between the virtual world and the physical world. As shown in \autoref{fig:demo}, it is a typical scene of IAR.

\subsection{Related Work}
In the past, a lot of related researches about augmented reality have been presented.
Before IAR, some novel styles of reality have been proposed. For example, Lifton \etal \cite{lifton2009dual} proposed the ``dual reality'' system to make the virtual world and the physical world be corresponding to each other. Roo \etal \cite{roo2017one} proposed the ``one reality'' system, which contained a 6-level mixture of virtual and real contents ranging from purely physical to purely virtual world. 
But they all describe the mixed reality from the perspective of humans, ignoring the view from the virtual world.

Since the virtual environment is expected to be the same intelligent as our natural environment in the physical world, it should be created with some virtual smart brains using current techniques in artificial intelligence.
Luck \etal \cite{luck2000applying} applied the artificial intelligence into virtual environments to make the virtual environments become intelligent. This work inspires us to add some intelligence to IAR, making the virtual world be driven by intelligence besides human's manual design. The intelligence of the virtual world can be accumulated by learning from the human's behaviors \cite{hattori2011learning} in some special cases. Note that the intelligence plays an important role in constructing IAR for the reason that the intelligence-driven self-development can make the virtual world act as the physical world does. If the intelligence is missing, the virtual world may suffer from low spontaneity, which can be harmful to the equivalence of the virtual world and the physical world. 

To make the virtual world intelligent, Taylor \etal \cite{taylor2017evolution} discussed the possibility of making a virtual world evolve by itself. The evolution of the virtual world took advantage of the principle of biological evolution in the physical world.  Though the self-learning is not simple, there are still many learning frameworks that can be used to obtain the self-learning ability, such as evolutionary computation \cite{fogel2006evolutionary}, reinforcement learning \cite{sutton1998reinforcement} and deep learning \cite{lecun2015deep}.

\subsection{Contribution}

In this paper, our main contributions are listed as follows. 
\begin{itemize}
\item Propose the concept of inverse augmented reality and elaborate the formulations according to physical properties.
\item Show the typical structure of inverse augmented reality systems and present the proof of concept for IAR.
\end{itemize}

\begin{figure}[tb]
 \centering % avoid the use of \begin{center}...\end{center} and use \centering instead (more compact)
 \includegraphics[width=\columnwidth]{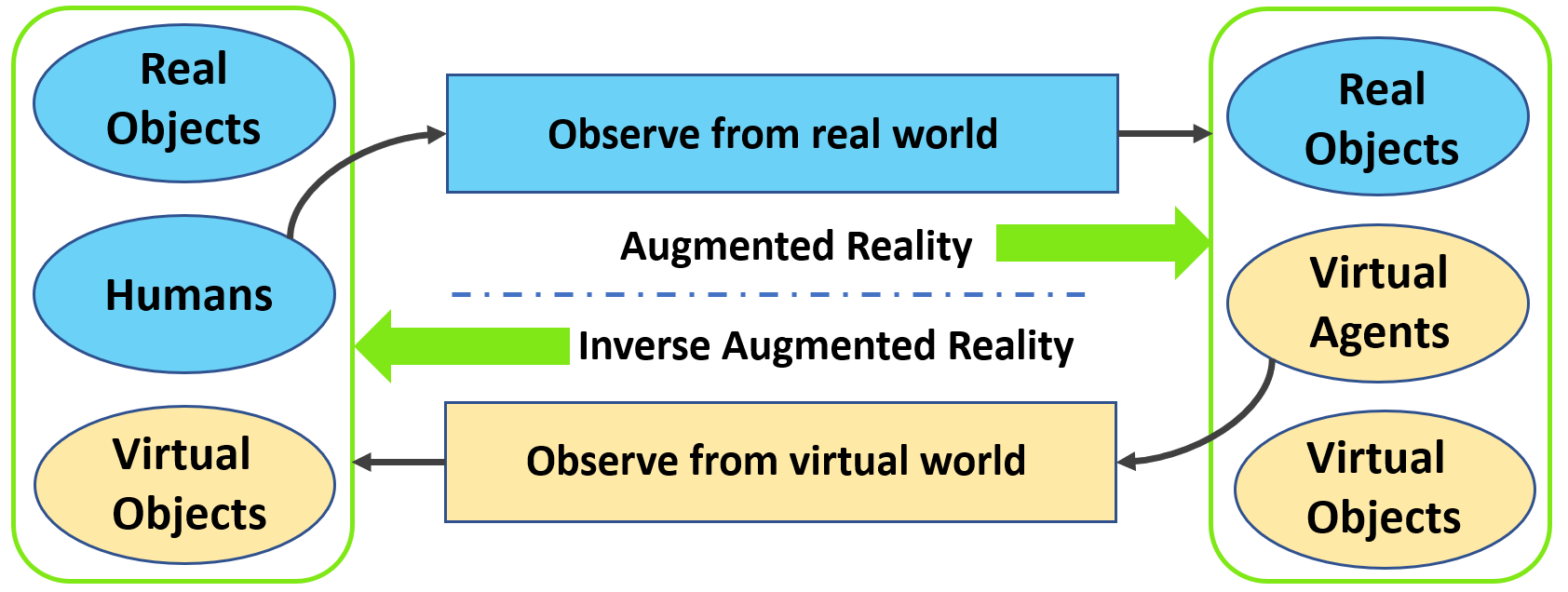}
 \caption{Relationship between AR and IAR. In the right rounded rectangle of the figure, it shows the typical scene of the traditional augmented reality, which can be observed by humans. In the left rounded rectangle of the figure, it shows the typical scene of the inverse augmented reality, which can be observed by virtual agents. Augmented reality and inverse augmented reality share the same structure except that they have opposite observing directions and different observers.}
 \label{fig:framework}
\end{figure}

\section{Framework of Inverse Augmented Reality} \label{sec:framework}

\subsection{Dual-World Structure }
The proposed inverse augmented reality and the traditional augmented reality, as shown in \autoref{fig:framework}, are under the unified dual-world structure. The traditional augmented reality (human-centered observation) is to augment the physical world with virtual objects, while the inverse augmented reality (virtual agent-centered observation) is to augment the virtual world with real objects. 

There might be a misconception between the proposed ``inverse augmented reality'' and another well-known concept called ``augmented virtuality''. Even though the two concepts are all describing using real elements in the physical world to augment virtual elements in the virtual world, their positions are definitely different. The augmented virtuality means that it is the human who can see a scene where the virtual elements are augmented by real elements, and the human himself is located in the real world. Conversely, the inverse augmented reality means that it is the virtual agent who can see a scene where the virtual elements are augmented by real elements, and the virtual agent itself is located in the virtual world.

\subsection{Mathematical Model}
Take the visual AR and IAR as the example, the formulation for AR and IAR can be as follows.

Let $O_R$ denote the real objects, $O_V$ the virtual objects, $H$ the humans, $A$ the virtual agents, then we get
\begin{equation}  \label{eq:AR}
\left\{  
    \begin{array}{lr}  
             AR \Leftrightarrow S_H(O_R, O_V, A)  \\  
             IAR \Leftrightarrow S_A(O_R, O_V, H)   
    \end{array}  
\right.
\end{equation}  

where $S_H$ denotes the observation function of humans, and $S_A$ denotes the observation function of virtual agents.

\section{Physical Perspective of Inverse Augmented Reality} \label{sec:IMR}
In this work, we emphasize the equivalence of the virtual world and the physical world regarding the structure in physics. The referred physics here contains both the physical world and the virtual world,\emph{ i.e.}, the virtual world is treated as a kind of existence in physics, which possesses the same structure with the physical world. In this way, IAR has the same important role as the traditional AR. We use a definition called physical equivalence to elaborate the equivalence of the physical world and the virtual world. This means the two worlds should be the same when talking about the physical structure, which can also be seen in \autoref{eq:AR}.

\subsection{Spatial Structure}
In the traditional augmented reality, there are three key components, \emph{i.e.}, the humans, the physical world and the virtual contents added to the physical world. As a correspondence, the same structure applies to inverse augmented reality. Concretely, inverse augmented reality also contains three key components, \emph{i.e.}, the virtual character, the programmable virtual world and the physical contents added to the virtual world. We emphasize the spatial structure rather than the appearance, because the difference regarding appearance is obvious. For example, all objects in the virtual world are data that are first created by human and then develop independently. Though the appearance is different, the spatial structure can be similar, especially the physical roles and interaction ways.

\subsection{Self Development}
As a common knowledge, the physical world we live in is keeping developing all the time. It seems to be driven by a kind of energy with the form of physical roles. Meanwhile, humans are born with intelligence, so they can actively interact with the physical world. Since the virtual world is expected to be developing by itself, it should have two kinds of agents, \emph{i.e.}, the character agent and the environment agent \cite{mateus2016intelligent}. The character agent can be treated as a virtual human in the virtual world, while the environment agent determines how the virtual environment can develop automatically. The two agents are created by our physical world, then they construct the virtual world and develop independently without being directly controlled by the physical world. The agents can not only learn from physical world but also evolve by themselves. Notice that only the character agents can observe things in the proposed framework of inverse augmented reality.

\subsection{Equal-Status Interaction}
\begin{figure}[tb]
 \centering 
 \includegraphics[width=\columnwidth]{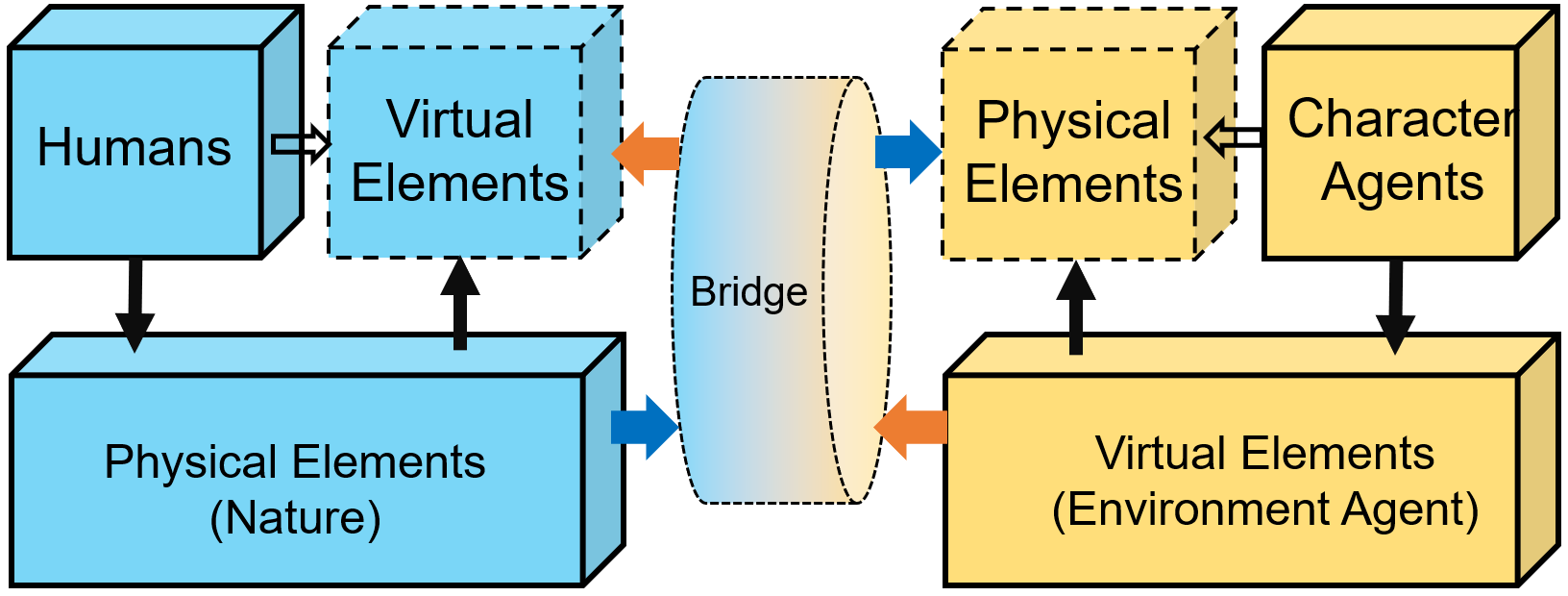}
 \caption{Interaction between physical world and virtual world. Solid black arrows denote the direct control stream. Hollow arrows denote indirect control stream. The big arrows with orange color denote the registration from virtual world to real world, while the big arrows with blue color denote the registration from physical world to virtual world.  }

 \label{fig:interaction}
\end{figure}

Considering the traditional AR and the proposed IAR, the physical world and the virtual world are equal to each other regarding interaction. As we often see in the traditional AR, a human can interact with both real and virtual objects that have been observed by him. Similarly, the character agent in the virtual world can interact with both virtual and real objects that have been observed by the agent. The two interaction processes are dual processes with the exactly symmetrical interaction style, as shown in \autoref{fig:interaction}. The interaction from virtual world to physical world means the virtual agent can control some physical power in order to change the physical state of real objects, \emph{e.g.}, if the virtual agent want to put a real box on a virtual table, it is required to find a certain physical way to support the real box so that it seems to be on the virtual table. And the physical way to realize this physical effect is expected to be controlled by the virtual agent. This is surely very hard for the current technology, but it is an essential part for IAR to support an equal interaction process compared with the traditional AR. Therefore, the equal-status interaction may need to be further studied and realized in the future. 

\section{Proof of Concepts} \label{sec:experiment}
\subsection{System setup}
We utilize Microsoft HoloLens as the basic platform to demonstrate the concept of IAR. Both AR and IAR are implemented by Unity3D and the Vuforia software development kit.

\subsection{Framework Representation}
\begin{figure}[tb]
 \centering 
 \includegraphics[width=\columnwidth]{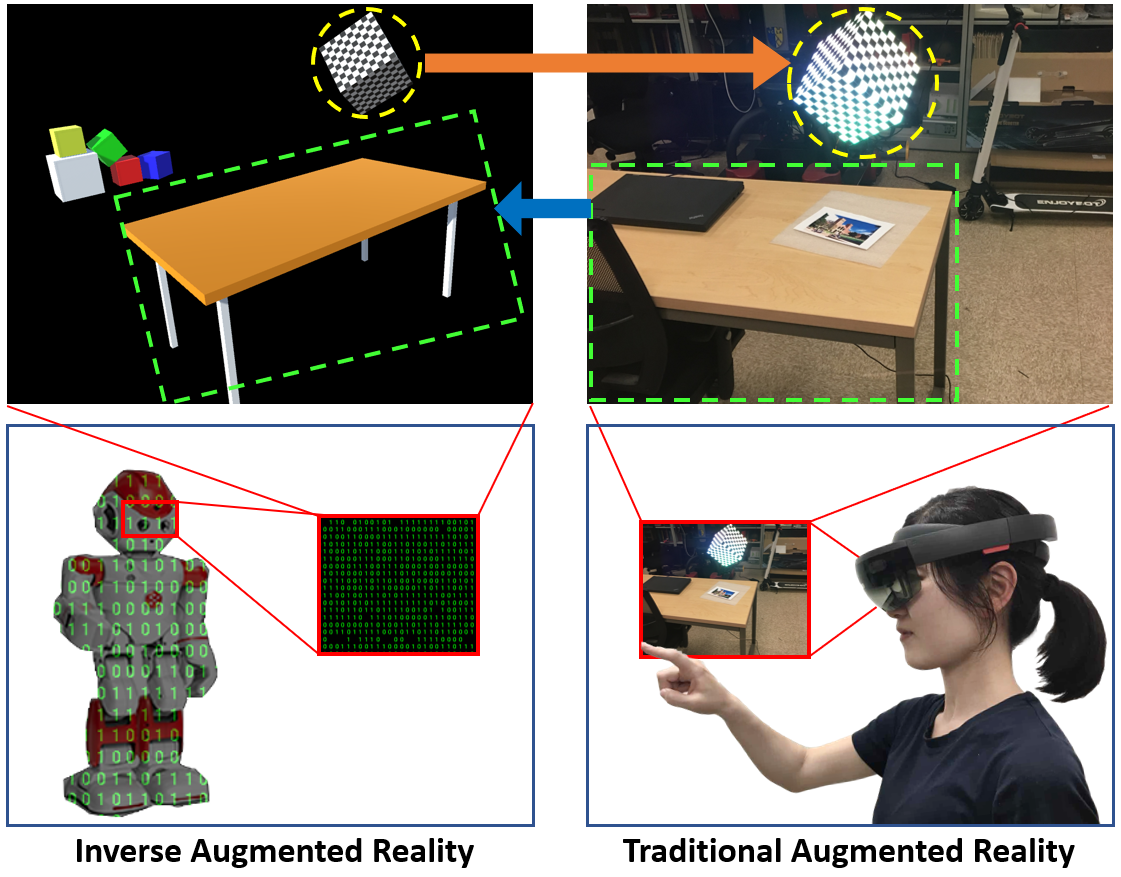}

 \caption{Concept demonstration. The left side is the view from the virtual world, while the right side is the view from the physical world. Some objects exist in both worlds, such as the cube with yellow dashed circle and the table with green dashed rectangle. Some objects only exist in one world. For example, some colorful cubes only exist in the virtual world, while a chair and a laptop only exist in the physical world.}
 \label{fig:exp_design}
\vspace{-3pt}
\end{figure}

Since the basic framework has been illustrated above, we present a typical demonstration of IAR using an office environment. We add a virtual cube floating above the table, which is located by a small photo. The small photo is fixed on the top of a table, which serves as a bridge connecting the physical world and the virtual world. After the environment is constructed, two views from the different worlds are shown in \autoref{fig:exp_design}. In the traditional augmented reality, the user can see the physical environment and the virtual element (a cube with the checkerboard pattern), and she can also interact with the virtual element. In the inverse augmented reality, a virtual agent is constructed, and it can behave like a physical human. Though what can be ``seen'' by the agent is absolutely some data, we can still figure out the meaning of these data. Usually, these data include the virtual cube that is connected with the physical world, the virtual table that corresponds to the real table in the physical world, and some other virtual objects that do not exist in the physical world.

\section{Discussion and Conclusion} 
The equivalence between the virtual world and the real world is proposed regarding the structure. As for the structure, it is already illustrated by introducing all essential parts of the traditional augmented reality and the inverse augmented reality. Though the specific expression forms are different, the two paradigms possess the same structure with each other. Our demonstration is about the concept verification, and all the results are shown directly by images observing from different worlds. This is a clear way to show the concept of IAR.

In this paper, we propose the big framework of the traditional augmented reality and the inverse augmented reality. Then we illustrate the main properties of this framework. Under this framework, we emphasize that the self-intelligence would play an important role in the virtual world, which contributes greatly to building an inverse augmented reality system. We also present a typical implementation of an inverse augmented reality system, which shows the inverse augmented reality can be realized with most current techniques.

The remaining challenges in the field of inverse augmented reality mainly include three aspects:

(1)	Physical construction of virtual objects in the physical world. 

(2)	Specific design of virtual-to-physical bridges.

(3)	Intelligence and knowledge for the self-driven virtual world.

Future work will be unifying the proposed IAR and the previous IVR into a more general framework in order to represent the reality at a higher level than what we have done currently. In this way, what the virtual agent could experience in both the virtual and the real world can be well illustrated.

%% if specified like this the section will be committed in review mode
\acknowledgments{
This work has been supported by the National Key R\&D Program of China (No. 2017YFB1002504) and the National Natural Science Foundation of China under Grant 61731003. }
\renewcommand{\baselinestretch}{0.92}
{
\bibliographystyle{abbrv-doi}

\bibliography{template.bib}
}
\end{document}